\documentstyle[12pt]{article}
\let\section=\subsection \let\subsection=\subsubsection

\newcommand{\beq}{\begin{equation}}
\newcommand{\eeq}{\end{equation}}

\newcommand{\bea}{\begin{eqnarray}}
\newcommand{\eea}{\end{eqnarray}}

         \makeatletter
         \def\thefigure{\@arabic\c@figure}\def\fps@figure{tbp}
         \def\ftype@figure{1}\def\ext@figure{lof}
         \def\fnum@figure{\protect\small Fig.\ \thefigure}
     \makeatother

\begin{document}
\begin{flushright}
{\rm MSUCL-1062}\\[9mm]
\end{flushright}
\begin{center}
{\large \bf DELAYS, UNSTABLE PARTICLES,}\\[2mm]
{\large \bf AND TRANSPORT THEORY\footnote{Talk given at the XXV
International Workshop
on Gross Properties of Nuclei and Nuclear
Excitations, ``QCD Phase Transitions'', Hirschegg,
January~13-18, 1997}
}\\[5mm]
P.~Danielewicz and S.~Pratt \\[5mm]
{\small \it
      National Superconducting Cyclotron Laboratory\\
       and Department of Physics and Astronomy\\
       Michigan State University, East Lansing, MI 48824-1321,
USA \\[8mm]}
\end{center}

\begin{abstract}\noindent
Delays associated with elementary interaction processes are
investigated.  The~case of broad resonances is discussed in the
context of reaction simulations.
\end{abstract}

\section{Introduction}
This study has been first motivated by problems within
the transport theory for low-energy heavy-ion reactions.
The~common microscopic models for the
reactions include the cascade and the Boltzmann-equation models.
In~the cascade model, a~reaction is represented as
a~superposition of independent NN collisions.  The
Boltzmann-equations models account for the mean-field effects,
in~addition to collisions, with particles moving along curved
rather than straight-line trajectories in-between the
collisions.  Within the models there is a~flexibility in
including different produced particles; the~mean-field
parameters may, principally, be~constrained using flow data.

\begin{figure}
  \vspace*{5.8cm}
  \includegraphics{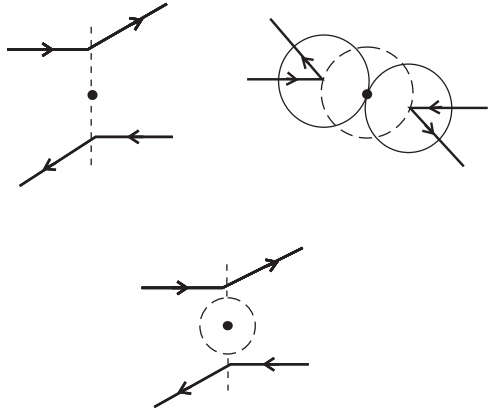}
\caption{
Different scattering prescriptions in transport
models: impact-parameter, billiard-ball, and mixed scattering.
}
\label{pres}
\end{figure}

In simulating the reactions, it is necessary to decide how to
simulate the elementary scattering processes.  Some
possibilities are indicated in~Fig.~\ref{pres}.
The~most common prescription is the one where particles change
their momenta at the distance of the closest approach.  This
constitutes the~so-called impact scattering.  Variations exist,
correlating the spatial separation between the particles and
the momentum transfer, as in the attractive (${\bf q r} < 0$)
and
repulsive (${\bf q r} > 0$) impact scattering.  At~low energies
cross sections are isotropic and the choices include
the~billiard-ball scattering.  Finally, the billiard-ball
scattering may be preferred at low impact parameters, and the
impact scattering at high.  This gives a~so-called mixed
scattering prescription.

While it is difficult to assess which scattering prescription
should be favored for interactions, the~results of reaction
simulations exhibit surprising sensitivity to the
prescriptions.  Thus, in~\cite{hal81}
variations by more than a~factor of~2 in the rise of density
above normal were found when using the billiard-ball and
the impact-parameter scattering for nucleons in a~central Ne +
U reaction.  In~\cite{gyu82} the
average flow angles were found to vary by nearly a~factor of~2
depending whether the impact or the repulsive impact scattering
was used in a U + U collision simulation.  Concerning high
energies,
Kahana {\it et al.} found huge flow effects in the simulations
of Au + Au collisions at 10.7~GeV/nucleon, when utilizing
the billiard-ball scattering.  Herrmann and
Bertsch~\cite{her95} considered
an~expanding pion gas in connection with the ultrarelativistic
collisions and obtained large differences in source radii, such
as studied through intensity interferometry, depending whether
pions followed attractive or repulsive impact scattering.

These and other experiences prompted an
investigation~\cite{dan96} to limit the short-term nature of
elementary processes and establish possible links between the
scattering prescriptions and the macroscopic properties of
a~system.

\section{Wave-Packet Interactions}

\begin{figure}
  \vspace*{4.3cm}
  \includegraphics{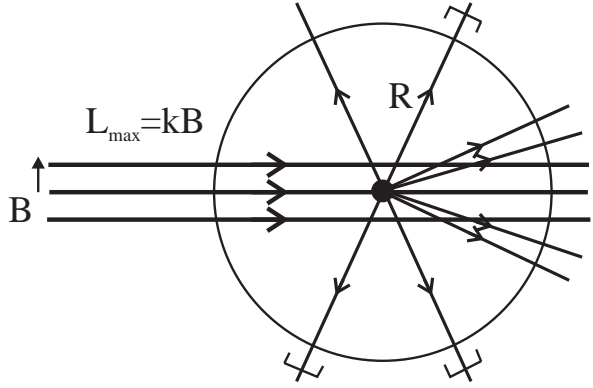}
\caption{
Scattering chamber of radius~$R$.
}
\label{chamber}
\end{figure}

To assess what happens in scattering we consider an~experiment
where a~target is located at the center of a~scattering chamber
of radius~$R$, see Fig.~\ref{chamber}.
A~wave-packet is directed,
towards the target, asymptotically
given by
\beq
\psi (t) = \int dE\ g(E)\ \left(
\phi^0 ({\bf r},t)  +
 f (\theta, E) \, {{\rm e}^{i(k r - Et)} \over r} \right)
\, ,
\eeq
where
\beq
\label{fi0}
\phi^0 = {1 \over 2i k r}
\sum_{\ell = 0}^{L_{max}} (2\ell+1) P_{\ell} (\cos{\theta})
\left[ (-1)^{(\ell +1)}
{\rm e}^{-i(k r + Et)} + {\rm e}^{i(k r - Et)} \right] \,
\eeq
and
\beq
f(\theta, E) =  {1\over k} \sum_{\ell} \, (2\ell + 1) \,
P_{\ell} (\cos{\theta}) \,
 \sin{\delta_{\ell}} \,
{\rm e}^{ i \delta_{\ell}} \, .
\eeq
The~wave-packet has a~transverse
size~$B$ and is constructed in such a~way that at the time
$t = 0$ it is centered at the target.

We examine the average
time spent by the packet within the chamber
\beq
\label{tau}
\tau_{vol} = R^2 \, {\int dt \, d\Omega \, {\bf
j} (R, \Omega, t) \cdot \hat{\bf r} \, t \over N} ,
\eeq
and compare it to the the time $\tau_{vol}^0 = 2R/v$ spent
within the chamber in the absence of interaction, $\tau_{vol}
= \tau_{vol}^0 + \Delta \tau$. In~(\ref{tau}),
${\bf j}$ is current density
\beq
\label{j}
{\bf j} = \frac{1}{2
\mu i} \, \phi^{*} \,
\stackrel{\leftrightarrow}{\partial} \, \phi
,
\eeq
and $N$ is integrated incident flux through the surface
of the chamber.  The~product
${\bf j} \cdot \hat{\bf r}$ in (\ref{tau}) is positive when
a particle exits the volume and negative when a~particle
enters.

For a large volume there is no interference between incoming
and outgoing waves
\beq
\tau_{vol} = \tau_{out} - \tau_{in} \, ,
\eeq
and the incoming time is found simply from (\ref{fi0}) as
$\tau_{in} = -R/v$.  The~outgoing waves consist of scattered
and unscattered portions of the wave packet and the
outgoing current has contributions from the scattered wave,
unscattered wave, and the interference between the two.

To~appreciate
the importance of interference one may consider the difference
between the net number of particles going out from and coming
into the chamber
\beq
 R^2 \, \int dt \, d\Omega \, {\bf
j} (R, \Omega, t) \cdot \hat{\bf r} = N_{out} - N_{in} \, .
\eeq
By integrating over the back of the chamber,
the~current coming in ($g(E) \simeq \delta (E) $) is found
equal to
\beq
\label{Nin}
N_{in} = { \pi \, v \over k^2} \sum_{\ell = 0}^{L_{max}}
(2\ell+1) = v \, \pi B^2 \, .
\eeq
The current going out consists of scattered and forward
parts,
\beq
N_{out} = N_s + N_f \, ,
\eeq
and
\beq
\label{Ns}
N_s = \int d \Omega \, {dN_s \over d \Omega} = \int
d \Omega \,
v | f |^2
= v \sigma \, .
\eeq
If the forward part is calculated from the unscattered packet
only, the~current coming out in the forward direction is
found to be the same
as that coming~in~(\ref{Nin}).  Thus, the~current coming out
through
wide angles~(\ref{Ns}) appears to be born within the chamber.
The~solution lies in a~careful treatment of the interference
between the unscattered portion of the packet and the scattered
wave.  In~(\ref{j}), the interference gives rise to terms
proportional to ${\rm e}^{ik(z-r)}$.  For finite angles the
terms vanish upon an~averaging over a~small macroscopic
range
of~$R$ but that is not the case in the forward direction.
On~integrating around the forward direction with the inclusion
of interference term, one finds
\beq
N_f =  { \pi \, v \over k^2} \left( \sum_{\ell =
0}^{L_{max}} (2\ell+1) - 4 k \, {\rm Im}
\, f (0) \right) \,
 = v \, \left( \pi B^2 - \sigma \right) = N_{in} - N_s \, .
\eeq
Due to $\sigma = {4 v \over k} \,  {\rm Im} \, f (0)$,
the~interference terms precisely correct for the flux moved
to finite angles.

Returning to the times, at finite angles there is no
interference and the average exit time may be computed as
a~function of scattering angle as
\beq
\tau_{s}(\theta) = R^2 \, \frac{ \int dt \,
{\bf j_{s}}(R,\theta,t) \cdot {\hat{\bf r}} \,
t}{dN_s/d\Omega}
 =   {R \over v} + {1
\over 2 i |f |^2 } \left\{ f^*
\, {d f \over d E} -
f \,
{d f^* \over d E} \right\}
 =  {R \over v} + {d \, {\rm Im} \, \log{f} \over dE} \, .
\eeq
The information on the delay $\Delta \tau_s$, relative to the
case without interaction, is contained in the {\em
phase} of the scattering amplitude.
For scattering in only one partial wave the delay time is
\beq
\Delta \tau_{s} =  \frac{d\delta_{\ell}}{dE} \, .
\eeq
Calculation of the average forward exit delay time yields
\bea
\nonumber
\tau_f & = & {R \over v} + {2
\over
\sum_{\ell}^{L_{max}} (2 \ell + 1) - 4 k \, {\rm Im}
\, f (0)} \,
{d \over d E}
\Big\lbrace k \, {\rm Re} \, f (0)
\Big\rbrace \\[.1cm] & \approx &
\frac{R}{v} + \frac{\sum_{\ell}^{L_{max}} \,
(2\ell+1) \, 2 \cos{2\delta_{\ell}} \,
\frac{d\delta_{\ell}}{dE}}{\sum_{\ell}^{L_{max}} \,
(2\ell+1) } \, .
\eea
The delay time in the forward direction involves the real
part of the forward scattering amplitude.

\section{Forward Delay and Mean Field}

One may ask whether any effect of the delays in elementary
processes is accounted for in the reaction simulations.  In~the
impulse approximation, the~real part of optical nuclear
potential may be expressed in terms of the real part of the
forward scattering amplitude,
\beq
U = n \, {\rm Re} \, {\cal T}(0) = - \, {N \over  V} \,
{2 \pi v \over k} \, {\rm Re} \, f (0) \, ,
\eeq
indicating a possible connection to the forward delay time.
Note that,
when density varies in space, so~does the optical potential.

\begin{figure}
  \vspace*{5.7cm}
  \includegraphics{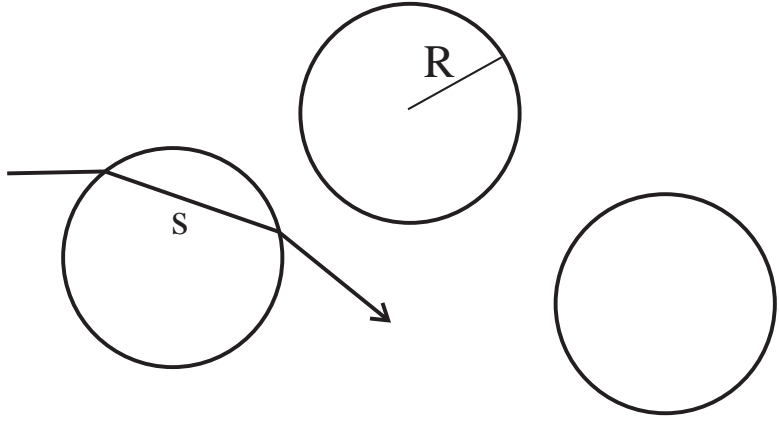}
\caption{
Particle incident on potential wells.
}
\label{scat}
\end{figure}

To further explore the possible connection between the optical
potential and the forward delay, let us consider a~situation
where we construct an~optical potential
landscape associated with some distribution of scatterers in
space.  One way to do it is to surround each scatterer with
a~well of radius~$R$ and depth ${\cal V}_0 = {\rm Re} \, {\cal
T}(0) / V$, where $V$ is the volume of a~well, and to
increase~$R$ until the wells merge, see~Fig.~\ref{scat}.
Let
a~particle be incident on a~set of the scatterers at
a~low density.  The~average time spent by the incident particle
within the region
of scatterers will be affected by the optical potential.
The~time of traversal through a~single well is $\tau
= s/v$, where $s$ is the length of trajectory within the well
and $v$ is group velocity.
When the optical potential is put to zero,
the~average
length of a~trajectory becomes just the ratio of the volume of
the
well to the transverse area, $\overline{s} = V/A = {4 \over 3}
\, \pi \, R^3 / \pi \, R^2 = {4 \over 3} \, R$.
The~average time within the well is
$\overline{\tau}
= \overline{s}/v$.
When
the potential is switched~on, the~time within the well
changes compared to the case without potential, due to
a~change in the length of trajectory caused by refraction,
and due to the change in the group velocity,
\bea
\nonumber
\Delta \overline{\tau} & = & { \Delta \overline{s} \over v} -
{\overline{s} \over v^2 } \, \Delta v = - { \overline{s} \over
V} \, {1 \over k^2} \,
{d \over d E}
\Big\lbrace {k^2 \over v} \, {\rm Re} \, {\cal T} (0)
\Big\rbrace  \\[.1cm]
& = & {2 \over k^2 R^2} \,
{d \over d E}
\Big\lbrace {k} \, {\rm Re} \, f (0)
\Big\rbrace
\approx {2 \over \sum_{\ell} \, (2 \ell +1) } \,
{d \over d E}
\Big\lbrace {k} \, {\rm Re} \, f (0)  \, .
\Big\rbrace
\eea
The~above manipulation shows that the change in the average
time for a~single
well is identical to the forward delay for a~packet of
transverse size equal to the size of the well, $\sum_{\ell} (2
\ell + 1) \approx k^2 R^2$.  Thus, the forward delay time
and the mean field are, indeed, directly related.

\section{Delays and Pressure}

Pressure for an~enclosed system may be defined as the momentum
delivered to the walls per unit area and per unit time.  Given
a~dependence of the pressure on density and temperature one can
determine other thermodynamic quantities for a~system.
In~the mean-field approximation, at~low density, the~correction
to the ideal-gas pressure from interactions can be, conversely,
obtained from the energy density.
Simple manipulations allow to express the
correction to the pressure in terms of the forward delay time,
\bea
\nonumber
P & = & P_0 + P_{mf} = n  \, T + {1 \over 2} \, n^2 \, {\rm Re}
\, {\cal T} (0) \\[.1cm]
& = & n \, T - {1 \over 2} \, T \int dp_1 \, dp_2 \, f_1 \, f_2
\, \pi R^2 \, v \, \Delta \tau_f  \, .
\eea
Given that particles stay shorter or longer within
the interaction region, compared to the free-flight time, they
knock more or less often against the enclosure walls and the
pressure increases or decreases, respectively, due to the
interactions.  Depending on the sign of the delay time the
correction due to the interactions may be further interpreted
in terms of excluded volume ($\Delta \tau < 0$) or a~reduction
in the number of degrees of freedom ($\Delta \tau > 0$).

Yet the delays associated with scattering should also affect
the frequency of knocking against the walls.  To~the
second-order in density the pressure is given completely by the
Beth-Uhlenbeck formula.  Manipulations of the second-order term
in the formula separate out the contributions to the pressure
from the forward delay time or mean field and from
scattering,
\bea
\nonumber
P & = & n \, T - T \, n^2  \, \left( {4 \pi \over
m T} \right)^{3/2} \, {1 \over 2} \int { dE \over 2 \pi} \,
{\rm e}^{-E/T}\,
 \sum_{\ell} \, (2\ell + 1) \, 2\, {d \delta_{\ell} \over dE}
\\[.1cm]
\nonumber
& = & P_0 + P_{mf} + P_{sc} = n \, T   -   {T \over 2} \int
dp_1 \, dp_2 \, f_1 \, f_2
\, \pi R^2 \, v \, \Delta \tau_f \\[.1cm]
& & \hspace*{1.55in}
 -  {T \over 2} \int
dp_1 \, dp_2 \, f_1 \, f_2
\, \int d \Omega \, {d \sigma \over d \Omega} \, v \,
\Delta \tau_s  \, .
\label{Pms}
\eea
The scattering can affect thermodynamic properties of a~system
just
as the mean field can.  Yet, the~nuclear equation of state in
connection with such reaction simulations as discussed in the
introduction is usually considered in terms of the mean field
alone.  It~is apparent from (\ref{Pms}) that, if~one aims at
describing solely properly the thermodynamic properties, then,
to~the second order in density, one can entirely absorb the
effects of scattering into the forward delay or the mean
field or, alternatively, the~effects of the mean field into
scattering.  The first is accomplished by using, in~place of
the forward delay~$\Delta \tau_f$, the~time
\beq
\Delta \tau_f'  =  \Delta \tau_f + { \pi R^2 \over \sigma} \,
\int d \Omega \, {d \sigma \over d \Omega} \, \Delta \tau_s \,
,
\eeq
or by making a replacement in the mean field
\bea
\nonumber
\lefteqn{
{\rm Re} \, {\cal T}(0)  =
- \, {\pi v \over k^2} \sum_{\ell} \, (2 \ell + 1) \sin{2
\delta_{\ell}}  } \hspace{5em}  & & \\
&
 \Rightarrow &
\hspace{1em} ({\rm Re} \, {\cal T} (0) )' =
- \, {\pi v \over k^2} \sum_{\ell} \, (2 \ell + 1) \, 2  \,
\delta_{\ell} \, .
\eea
The effects of the mean field may be absorbed into scattering
by using, in~place of~$\Delta \tau_s$,
the~time
\beq
\Delta \tau_s'  =  \Delta \tau_s + {\sigma \over \pi R^2} \,
\Delta \tau_f \, .
\label{tausp}
\eeq

At a~general level, the relation of thermodynamic properties to
the delays may be understood in terms of ergodicity.
The~system spends time in a~certain phase-space
region, that is proportional to the density of states in that
region.  Changed time corresponds to the changed density of
states and thus to changed thermodynamic properties.

\subsection{Resonance Interactions}

\begin{figure}[tb]
  \vspace*{6.2cm}
  \includegraphics{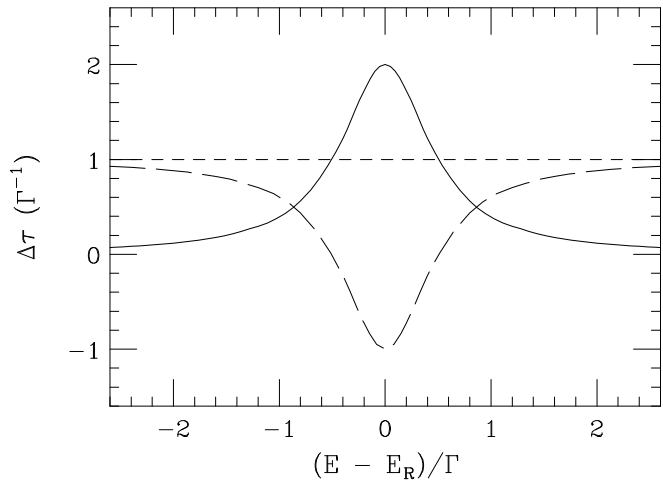}
\caption{
Time delays in the case of a~Breit-Wigner resonance as a
function
of energy from the resonance divided by the width.  The solid line
represents the time for the scattered wave $\Delta \tau_s$.
The short-dashed line represents the time $\Delta \tau_s '$
which
is ergodically consistent when the delay for forward wave or
the mean
field are neglected.  Finally, the long-dashed line shows the
time delay for the forward wave divided by the fraction of the
incoming wave that is scattered, $\Delta \tau_s '-\Delta
\tau_s$.
}
\label{BW}
\end{figure}

One of the important examples of interactions to consider is
that of a~Breit-Wigner resonance.
The~phase shift for the
resonance is given by
\bea
\tan{\delta} = - \frac{\Gamma /2}{E-E_R} \, .
\label{tand}
\eea
In transport simulations the resonances are given, intuitively,
a~lifetime~$\Delta \tau_s = 1/ \Gamma$.  From~(\ref{tand}) the
actual scattering delay is
\beq
\Delta \tau_{s} = \frac{d\delta}{dE} =
\frac{\Gamma} {2[(E-E_R)^2 + \Gamma^2/4]} \, .
\label{taus}
\eeq
At $E=E_R$, the~scattering time is twice as large as the
naively expected and it tends to zero as $|E - E_R|$
gets large, see~Fig.~\ref{BW}.
The forward delay time, normalized in such a way as if it were
to be put into scattering, is
\beq
\Delta \tau_f \times {\pi B^2 \over \sigma}  = \Delta \tau_s' -
\Delta \tau_s = {(E - E_R)^2 - \Gamma^2/4
\over \Gamma [(E - E_R)^2 + \Gamma^2/4 ] } \, .
\label{tauf}
\eeq
This time is negative in the vicinity of resonance.  That
corresponds, in~particular, to the known increase in the group
velocity for light near resonances in dielectrics.

The sum of the times (\ref{taus}) and (\ref{tauf}) is $\Delta
\tau_s' = 1 / \Gamma$.  Thus the naive scattering time accounts
for {\em both} the scattering and the mean field.  When
employing such
a~time in a~simulation it is inappropriate to include
separately the effect of the specific resonance onto mean
field.  For narrow resonances, under averaging over energy,
the~time $\Delta \tau_s$ averages itself out to $1 /
\Gamma$,
as~the mean field averages out to zero.  However, the~averaging
would not be justified for broad resonances.

\begin{figure}
  \vspace*{6.2cm}
  \includegraphics{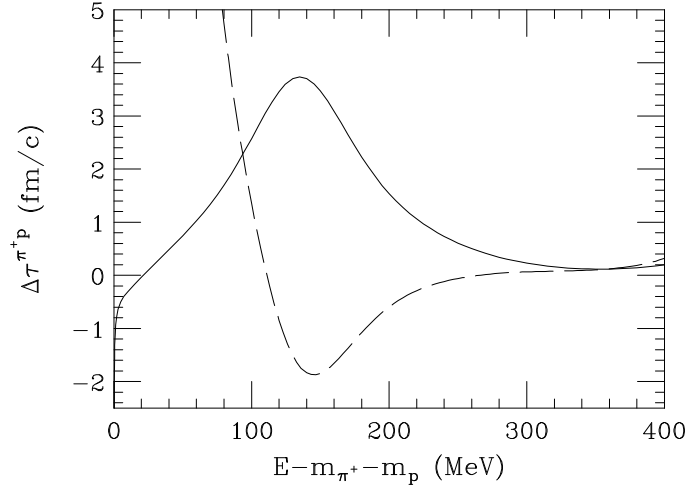}
\caption{
Time delays for a~$\pi^+ p$ system as a function of
c.m.~kinetic energy, computed using measured phase shifts.
The~solid line represents the time delay for
the scattered wave averaged over angles and spin directions,
$\Delta \tau_s$.  The~dashed line represents the forward time
delay
averaged over spin directions and divided by the fraction of
the incoming wave that is, on the average, scattered,
$\Delta \tau_s '-\Delta \tau_s$.
}
\label{delta}
\end{figure}

One of the more important broad resonances in high-energy
nuclear physics is the $\pi N$ $\Delta$-resonance.  The~width
for this resonance
is comparable to the energy above the threshold and, as such,
the width exhibits a~strong energy dependence.  In~the
immediate vicinity of the threshold, the~width behaves~as
\bea
\Gamma \propto (E - m_{\pi} - m_N)^{3/2}  \, ,
\eea
The~use of the time $1 / \Gamma(E)$ in simulations leads to
very long-lived $\Delta$'s close to the threshold, which
must be unphysical.  Intuitively,
the $\Delta$'s far from the center of the resonance
are expected to be virtual.  Figure~\ref{delta}
shows the delay
times $\Delta \tau_s$ and $\Delta \tau_s' - \Delta \tau_s$
computed from the measured $\pi N$ phase shifts.
It is apparent that the
scattering delay time $\Delta \tau_s$ has a~quite regular
behavior with energy.  It~is $\Delta \tau_s ' - \Delta \tau_s$,
within $\Delta \tau_s' \sim 1 / \Gamma$, that strongly diverges
close to the threshold for the $\Delta$ resonance production.
It then follows that
the physically unacceptable  divergence of $\Delta
\tau_s'$ in simulations is associated with forcing the mean
field effects onto scattering when actually the~scattering
cross section declines, $\sigma_{\pi N \rightarrow \Delta}
\propto (E - m_{\pi} - M_N)^{1/2}$.

With the above, the solution to the dilemma with diverging
times appears straightforward.  The~scattering and forward
delays
should simply be treated separately.  Yet the delays in
a~simulation
accomplish various goals.  One goal is the production of the
proper
overall density of states and of the corresponding proper
thermodynamic properties in equilibrium.  Another goal is the
production of the proper number of resonances which, when
decaying,
may contribute to the energetic photons or lepton pairs,~etc.
In~equilibrium, the~number of $\Delta$ resonances is given in
terms of the resonance spectral function and the use of the
time $\Delta \tau_s' = 1/ \Gamma$ in scattering, during which
a~$\pi N$ pair converts into a~$\Delta$,
meets the requirement of the proper number of resonances, see
below.  One may expect that, when replacing
$\Delta \tau_s'$ by $\Delta
\tau_s$ and $\Delta \tau_f$, and when converting $\pi N$ pairs
into the $\Delta$'s during these times, the~requirement will
still be met,
\bea
\nonumber
n_\Delta (\mu,T) & = & 16 \int d{\bf P} \, dm \,
{\rm e}^{(\mu - E)/T} \,
{\Gamma \over (m - m_\Delta)^2 + \Gamma^2/4} \\[.15cm]
\nonumber
& = &
12 \int d{\bf P} \, d{\bf p} \,
{\rm e}^{(\mu - E)/T} \,
\, v \, \, {16 \over 12} \, { \pi \over p^{*2}}
\, { \Gamma^2 \over (m - m_\Delta)^2 + \Gamma^2/4} \, \, {1 \over
\Gamma} \, \, {E \over m}
\\[.15cm]
& = &
12 \int d{\bf P} \, d{\bf p} \,
{\rm e}^{(\mu - E)/T} \,
\, v \, \sigma_{N \pi \rightarrow \Delta} \, {\gamma \over
\Gamma} \, \, \,
\stackrel{\textstyle ?}{=} \, \, \,
\mbox{scat} + \mbox{for}  \, .
\eea
However, the problem which arises, is that the
forward delay
time for a~resonance~(\ref{tauf}) is not positive definite, and
one cannot simulate a~negative time for the conversion of
a~$\pi N$ pair into a~$\Delta$.  A~modification, relying on the
possibility of moving the mean-field effects into scattering
and back, can render the scattering and forward times
both positive and
free from any singularity, e.g.~with
\beq
\Delta \tau_{s \Delta} = { \Gamma/4 \over (m - m_\Delta)^2 +
\Gamma^2/4} \hspace{3em} \Delta \tau_{f \Delta} = { \sigma
\over
\pi B^2} \, {(m - m_\Delta)^2  \over \Gamma [ (m -
m_\Delta)^2 + \Gamma^2/4 ] } \, .
\label{tdelta}
\eeq
For the forward processes the $\Delta$ should decay back into
the original pion and nucleon with their original momenta.
The~times (\ref{tdelta}) are consistent with the ergodic
constraint, up to terms proportional to $\partial \Gamma /
\partial m$ and $\partial m_\Delta / \partial m$, and with the
proper number of $\Delta$'s, as~related to the number of $\pi N$
pairs.

\section{Schemes in Simulations}

The sensitivity of the reaction simulation to prescriptions in
scattering, discussed in the introduction, can be understood in
terms of scattering
delays and changes made to the nuclear equation of state.
The~billiard-ball
scattering gives an average delay time equal to the negative of
the time within hard-core in a~free passage,
\beq
\Delta \tau_s = - {1 \over v} \, {{4 \over 3} \pi d^3 \over \pi
d^2} = - {4 \over 3} \, {d \over v} \, .
\label{bill}
\eeq
The~insertion of this time delay into
the pressure~(\ref{Pms}) yields the repulsive excluded volume
correction
to the pressure, valid to within the second order in density,
\beq
P = nT - {1 \over 2} \, n^2 \, T \, \langle \sigma \, v \,
\Delta
\tau_s \rangle =  nT + n^2 \, T \,  {2 \pi d^3 \over 3} \,  .
\eeq
The sign of the time delay and of the correction to pressure
may be inverted, making the correction to pressure attractive,
by replacing the hard core in the scattering by a~spherical
shell of radius~$d$ open in the direction of motion,
(essentially replacing the convex by the concave mirror).

The repulsive impact-parameter scattering yields the average
time delay and the correction to pressure nearly as large as
the billiard-ball scattering,
\beq
\Delta \tau_s = - {\pi \over 3} \, {d
\over v} \, ,
\label{impact}
\eeq
i.e.\ $\pi /4$ of~(\ref{bill}).
The~attractive impact-scattering changes the sign in
(\ref{impact}) and in the~correction to pressure.  Finally, for
the impact scattering the average time-delay is zero and the
correction to pressure from scattering vanishes, within the
second order in density.
Given {\em required} delay times obtained from
phase shifts~\cite{dan96}, such that $|\Delta \tau_s| < (2/3)
\, d \, \langle \sin{\theta} \rangle = (\Delta \tau_s)_{max}$,
these times may be generated making a~fraction $\nu$ of all
scatterings repulsive and a~fraction $1 - \nu$ attractive,
\beq
\nu = {1 \over 2} \left( 1 - {\Delta \tau_s \over (\Delta
\tau_s )_{max} } \right) \, .
\eeq

\section{Conclusions}

The two- or more-body processes hardly ever take place
instantaneously.  In~analyzing the wave-packet scattering we
found that, in the consequence of interaction, both the
forward
and the scattered waves get delayed.  The~forward delays are
accounted for in the mean field in reaction simulations.
Both types of delay, though, through ergodicity, affect the
density of states in a~system and
the~system thermodynamic
properties.  To~the extent that only the density of states
matters, then, to~the lowest order in spatial density,
the~same density of states and the same thermodynamic
properties may
be obtained by forcing all delays onto scattering or onto the
forward direction.  Some difficulties, though, may arise in
specific situations. Thus, the~unphysically long lifetimes of
resonances
close to threshold in simulations are specifically associated
with the attempt to force the mean-field effects onto
scattering.\\[-1ex]

This work was partially supported by the National Science
Foundation under Grant PHY-9403666.

\end{document}